\documentclass[]{elsarticle}

\usepackage[square]{natbib}

\usepackage[dvips,pdftex]{graphicx} 
\usepackage{subfigure} 
\usepackage{float} 

\usepackage[latin1]{inputenc}
\usepackage[T1]{fontenc}

\usepackage{amsmath}	
\usepackage{amssymb}  
\usepackage{amsbsy}
\usepackage{amsfonts}
\usepackage[tight]{units} 
\usepackage{textcomp}

\newcommand{\beq}{\begin{eqnarray}}
\newcommand{\eeq}{\end{eqnarray}}

\begin{document}

\begin{frontmatter}

\title{Position-sensitive ion detection in precision Penning trap mass spectrometry}

\author[]{G. Eitel$^{\textnormal{a,}}$\footnote[1]{present address: Aerodynamisches Institut, RWTH Aachen University, W\"ullnerstrasse 5a, 52062 Aachen, Germany\\Tel.: +49 (0)241 80 90400\\Fax: +49 (0)241 80 92257\\Email: g.eitel@aia.rwth-aachen.de}$^,$\footnote[2]{The present work deals with results partly presented in the diploma theses of G.~Eitel and J.~Ketter.}}
\author[label2]{M. Block}
\author[label3]{A. Czasch}
\author[label2]{M. Dworschak}
\author[label1,label4]{S. George}
\author[label3]{\mbox{O. Jagutzki}}
\author[label1]{J. Ketelaer}
\author[]{J. Ketter$^{\textnormal{a,}}$\footnotemark[2]}
\author[label2,label4]{Sz. Nagy}
\author[label5]{D. Rodr\'iguez}
\author[label6,label7]{C.~Smorra}
\author[label4,label7]{K. Blaum}

\address[label1]{Institut f\"ur Physik, Johannes Gutenberg-Universit\"at Mainz, 55099 Mainz, Germany}
\address[label2]{GSI Helmholtzzentrum f\"ur Schwerionenforschung mbH, 64291 Darmstadt, Germany}
\address[label3]{Institut f\"ur Kernphysik, Johann Wolfgang Goethe-Universit\"at Frankfurt am Main, \mbox{60438 Frankfurt am Main, Germany}}
\address[label4]{Max-Planck-Institut f\"ur Kernphysik, 69117 Heidelberg, Germany}
\address[label5]{Departamento de F\'isica At\'omica Molecular y Nuclear, Universidad de Granada, \mbox{18071 Granada}, Spain}
\address[label6]{Institut f\"ur Kernchemie, Johannes Gutenberg-Universit\"at Mainz, 55099 Mainz, Germany}
\address[label7]{Ruprecht-Karls-Universit\"at Heidelberg, 69120 Heidelberg, Germany}

\begin{abstract}
A commercial, position-sensitive ion detector was used for the first time for the \mbox{time-of-flight} ion-cyclotron resonance detection technique in Penning trap mass spectrometry. In this work, the characteristics of the detector and its implementation in a Penning trap mass spectrometer will be presented.\newline In addition, simulations and experimental studies concerning the observation of ions ejected from a Penning trap are described. This will allow for a precise monitoring of the state of ion motion in the trap.
\end{abstract}


\end{frontmatter}

\section{Introduction}
\label{sec:introduction}
In recent years, Penning trap mass spectrometry has proved to be the best technique for high-precision mass measurements on short-lived and stable nuclides~\cite{Blaum:06}. Here, ions with a charge-to-mass ratio $q/m$ are confined in a strong homogeneous magnetic field $B$ superposed with a weak electrostatic quadrupole field~\cite{Brown:86}. The mass measurement is carried out via the determination of the characteristic cyclotron frequency $\nu_c=qB/(2\pi m)$. In the case of radionuclides, mass uncertainties below $10^{-8}$~\cite{Blaum:03b,Kellerbauer:04a,Bollen2006,george2007PRL,Hager2006,Ryjkov2008,rauth:08,savard:05} have been reached.
The mass measurement is usually performed with the des\-truc\-tive \mbox{time-of-flight} ion-cyclotron resonance \mbox{(TOF-ICR)} detection~\cite{grae80} suitable for very short-lived radionuclides with half lives $T_{1/2}<1\,s$. This technique was introduced two decades ago and it has been continuously improved. However, until now only information about the number of ions and their time of flight is obtained typically by an ion counting detector and a time digitizer. A detector with high ion detection efficiency is essential, such as a micro-channel plate (MCP) detector~\cite{wiza79} where in a Chevron configuration with an adequate bias network detection efficiencies of~$30-50$\,$\%$ are being reported~\cite{Starub:99}. Another electron multiplier, called Channeltron~\cite{yazidjian2006ncd}, can reach absolute detection efficiencies of about~90\,\%.\newline
In on-line mass measurements using radioactive nuclides the presence of contamination has to be expected. The contaminants can be isobaric or isomeric impurities from the source itself, decay products of short-lived radioactive ions, or produced in the experimental setup by charge exchange reactions with the buffer-gas or rest-gas atoms. Depending on the number of contaminant ions the mean time of flight is altered and the measured cyclotron frequency is shifted to lower values. The effects of the presence of contamination in the precision trap have been described in detail in~\cite{boll92}.\newline
In this article an improvement of the TOF-ICR method using a high resolution position-sensitive MCP detector is introduced. The detector offers a tool to precisely monitor the state of ion motion in the trap. Furthermore, the \mbox{signal-to-noise} ratio of the observed \mbox{time-of-flight} resonance can be improved by excluding non-excited contaminant ion species which are identified by their position on the detector. Extensive ion trajectory simulations were performed to calculate the impact position of the excited ions after ejection from the Penning trap.\newline
In the following, the classical ion motion in an ideal Penning trap and the ion excitation techniques relevant to this work are briefly reviewed. After that, simulations and experimental results using the new detection technique are discussed.

\section{Ion motion and excitation in a Penning trap}
The superposition of a strong homogeneous magnetic field $B$ and a weak electrostatic field, called Penning configuration, allows for three-dimensional confinement of charged particles. The ion motion in a Penning trap is well understood, and the classical equations of motion are easily derived~\cite{Blaum:06,Brown:86}.
\begin{figure}[htpb]
  \centering
    \includegraphics[width=0.8\textwidth]{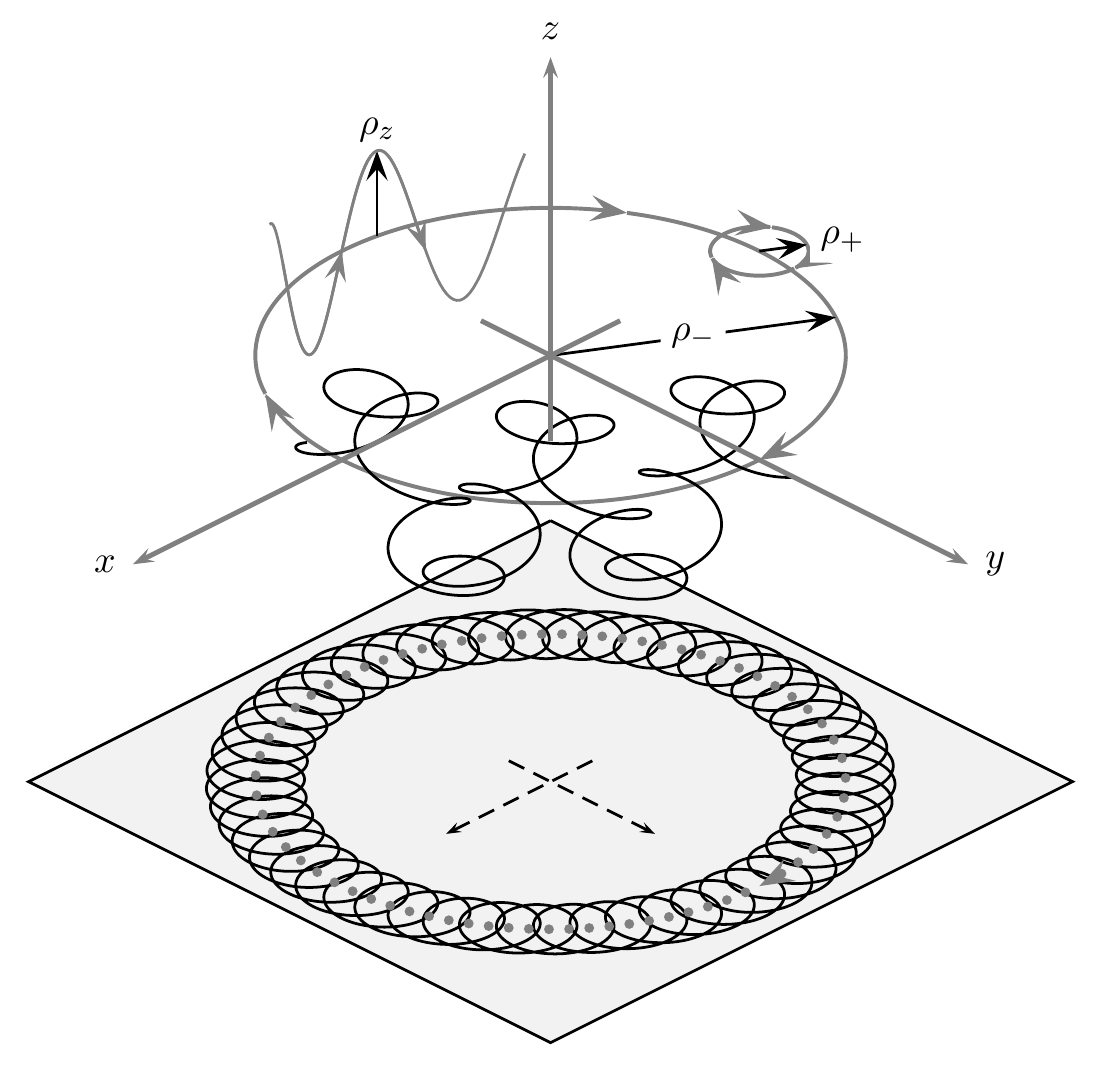}
    \caption{Ion motion in an ideal Penning trap. $\rho_-$ is the radius of the magnetron motion, $\rho_z$ is the amplitude of the axial oscillation, and $\rho_+$ is the radius of the cyclotron motion. The superposition of the three eigenmotions is described by a black solid line. In the ($x,y$) plane the projection of the three-dimensional motion is shown for a complete magnetron period. The frequencies and amplitudes are not to scale.}
    \label{fig:theory_ion_motion}
\end{figure}
The motion of the trapped particle is a superposition of three independent oscillations: One oscillatory motion in the axial-direction with the frequency $\omega_z$, 
and two orbital motions in the radial plane: the slow oscillatory magnetron motion with the eigenfrequency $\omega_-$, and the fast oscillating cyclotron motion with the eigenfrequency $\omega_+$. All three eigenmotions are illustrated in Fig.\,\ref{fig:theory_ion_motion}. A series expansion of the radial motions shows that the magnetron frequency is mass independent to first order. The hierarchy of the frequencies is $\omega_-<\omega_z<\omega_+<\omega_c$ and important relations between the eigenfrequencies with respect to mass spectrometry are
\begin{equation}
\omega_c = \omega_+ + \omega_{-}
\label{eq:fsum}
\end{equation}
\begin{equation}
\omega_c^2 = \omega_{+}^2 + \omega_{z}^2 +\omega_{-}^2\,.
\end{equation}
The latter is known as the \emph{invariance theorem}~\cite{Brown:82} and is even valid for a Penning trap with small imperfections in the alignment of the electric and magnetic field~\cite{gabrielse2009}.\newline
In the ideal case the three oscillation modes are uncoupled and each can be described by a
quantized harmonic oscillator~\cite{Kret:92}. An external time dependent electric field can 
be used to manipulate the amplitude of certain eigenmotions of the stored ions. Therefore, usually two different driving fields are employed, a dipole and a quadrupole field.

\subsection{Dipolar excitation}
Dipolar excitation at a specific eigenfrequency allows the manipulation of the amplitude of the respective eigenmotion. By applying a dipolar excitation at the mass independent frequency $\omega_-$ it is possible to change the magnetron amplitude of all trapped ion species simultaneously. Mass-selective dipolar excitation at the modified cyclotron frequency $\omega_+$ is used to remove unwanted ion species from the trap. A detailed discussion can be found in~\cite{Blaum:03}.

\subsection{Quadrupolar excitation and time-of-flight ion-cyclotron resonance detection}
A quadrupolar driving field at a frequency $\nu_{rf}$ equal to the sum of two individual eigenfrequencies, e.g. $\nu_{rf} = \nu_c = \nu_+ + \nu_-$, leads to a coupling of the corresponding eigenmotions, which results in a periodic interconversion between the two modes~\cite{bollen1990,koen95}. A full conversion is obtained after a time $T_{conv}$ which is half the beating period. In Penning trap mass spectrometry a quadrupolar excitation is applied for mass-selective buffer-gas cooling and for the determination of the cyclotron frequency $\nu_c$ by means of the TOF-ICR technique. Thereby the stored ions are probed with a radio frequency $\nu_{rf}$. Thus the ion's radial energy is increased to a maximum in case that $\nu_{rf}$ matches the cyclotron frequency $\nu_c$. When the ions are ejected from the trap the interaction of the magnetic moment of the ion motion with the gradient of the magnetic field causes an axial acceleration, i.e. a conversion of radial energy to axial energy takes place. Thus, the time of flight from the trap to an ion counting detector reaches a minimum if the previous excitation was resonant. To calibrate the $B$-field an ion with well-known mass is used and the mass of interest is obtained from a cyclotron frequency ratio~\cite{kell03}.

\section{Description of the delay-line detector}
\label{sec:Detector} 
For the present work, a position-sensitive MCP detector (DLD40) from \mbox{RoentDek GmbH}~\cite{jagutzki2002bam,roent} was used. The DLD40 is a combination of an MCP in Chevron configuration and a so-called delay-line anode \cite{lampton1987dla}. The anode consists of two wire pairs coiled in $x$- and $y$-directions, which are set to a positive potential of about $\unit[300]{V}$ with respect to the MCP output. An incident charged particle triggers an avalanche of secondary electrons in one of the MCP's channels. These electrons induce an electric signal on the anode wire, which propagates along the delay-line wires. The two-dimensional position is encoded by the difference in signal propagation time to each end of the delay-line. This timing information is determined with a time-to-digital converter (TDC) with a least significant bit of $\unit[25]{ps}$. This results in a spatial resolution for a single event of $\unit[70]{\mu m}$ in both dimensions~\cite{roent:DLDman}. The complete imaging equipment includes the DLD40 detector unit, a high-voltage supply (NHQ214M), an amplifier/CFD unit with NIM output format (ATR19), a TDC PCI-card (TDC8HP), and software for data acquisition. In Tab.\,\ref{tab:MCPparameters} the elementary technical specifications of the detector are listed, including a value for the absolute detection efficiency of $\unit[30-40]{\,\%}$ relating to ions of $^{85}$Rb and $^{133}$Cs with low kinetic energy. For high-energy alpha particles a slightly higher efficiency of $\approx\unit[60]{\%}$ has been determined. Usually the provided read-out software is designed for permanent data-acquisition. In order to allow external triggering, the read-out software was modified in close collaboration with the manufacturer. These modifications included consistency conditions, which ensured a correct time check-sum and the presence of all four anode signals. The acquired data is recorded in a list-mode file (LMF) with a set of five timing values for each event. By using a graphical user interface, the data can be visualized during and after the acquisition in arbitrary plot styles and the LMF can be processed further.\newline
Due to an active diameter of more than $\unit[42]{mm}$ and its high spatial resolution, the detector is suited to serve for precision ion beam diagnostics. The employed detection principle tolerates no environmental magnetic field strength above several mT. A follow-up model, capable of stronger magnetic fields, is being developed by RoentDek GmbH~\cite{jagutzki2007} and has been tested recently at the \mbox{Max-Planck-Institute} for Nuclear Physics in Heidelberg. In the following, the DLD40 detector is employed as a tool to investigate the state of ion motion in a Penning trap.
\begin{table}[H]
    \centering
        \begin{tabular}{|l|c|}
            \hline
            Outer Diameter & $\unit[50]{mm}$ \\
            Active Diameter & $\unit[>42]{mm}$ \\
            Channel Diameter & $\unit[25]{\mu m}$ \\
            Bias Angle & $7^\circ\pm 2^\circ\ $ \\
            Operating Voltage (2 MCPs) & $\unit[-2400]{V}$ \\
            Anode Voltage & $\unit[+300]{V}$ \\
            Rate Capability & $\unit[10^6]{Events/s}$ \\
            Spatial Resolution & $\unit[70]{\mu m}$ \\
            Detection Efficiency & $\unit[30-40]{\%}$ \\
            Pulse Width & $\approx\unit[9]{ns}$ \\
            Signal Rise Time& $\approx\unit[1.5]{ns}$ \\
            \hline
        \end{tabular}
    \caption[Specifications of the Delay-Line Detector]{Specifications of the Delay-Line Detector. The presented values refer to measurements with singly charged positive ions.}
    \label{tab:MCPparameters}
\end{table}

\section{Computational simulations}
\label{sec:simulation}
\begin{figure}[htpb]
  \centering
    \includegraphics[width=0.8\textwidth]{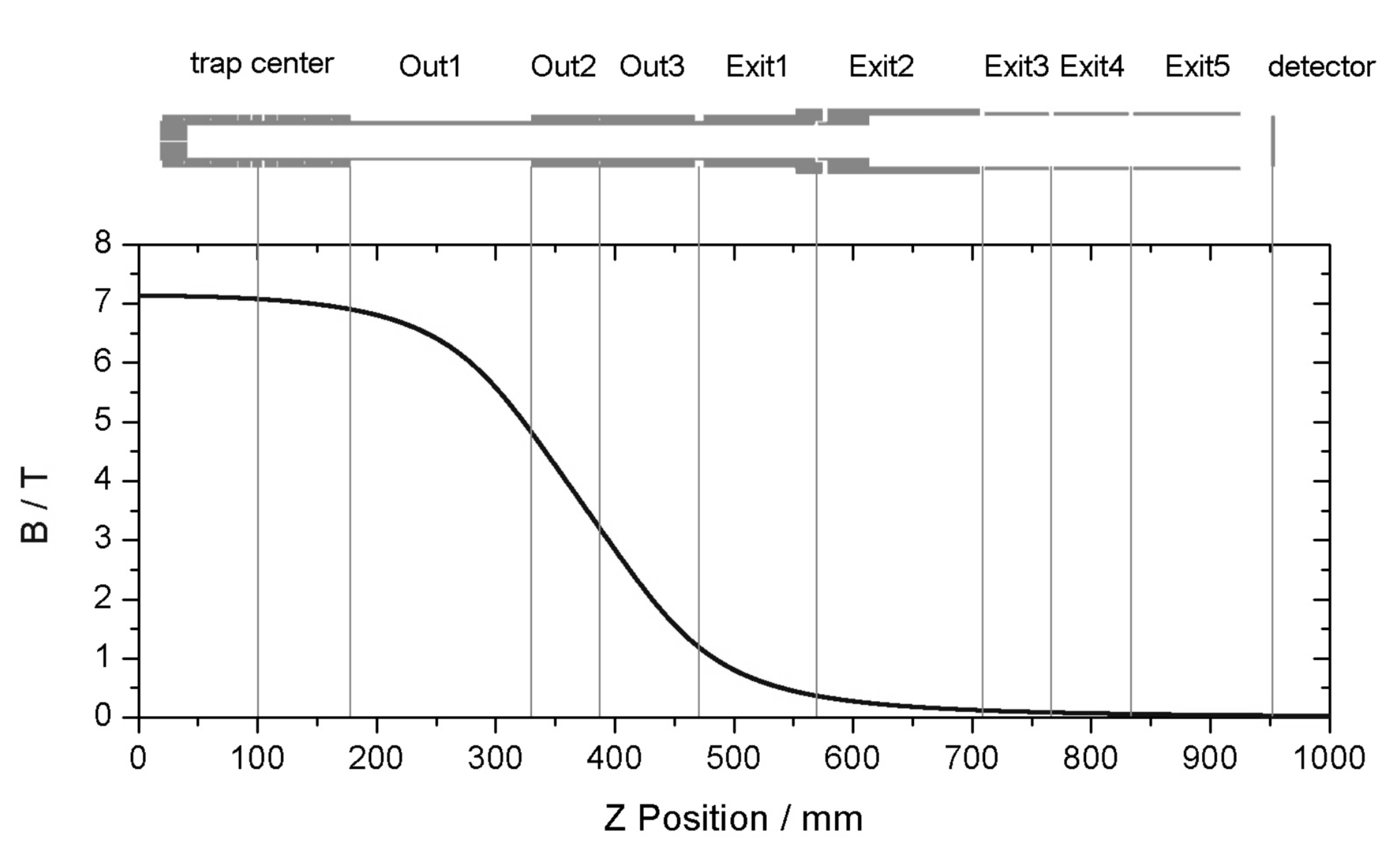}
    \caption[]{Sections of the drift tube and the magnetic field strength along the $z$-axis according to the SHIPTRAP mass spectrometer.}
    \label{fig:simulation_fieldstrength}
\end{figure}

The simulation studies presented here address the trajectories of single ions ejected from a Penning trap. The initial conditions of the ion motion were computed by using the analytical treatment of the excitation of radial eigenmotions in an ideal Penning trap~\cite{koen95}. For the simulation of the ejection sequence, the commercial software \mbox{SIMION 3D\textregistered~8.0}~\cite{simi80} was used. SIMION 3D\textregistered~is a tool for the numerical calculation of ion trajectories in a given electrode geometry~\cite{dahl2000}. The trap setup simulated here includes a cylindrical measurement trap, a drift section, and a magnetic field, according to the SHIPTRAP setup~\cite{bloc05} briefly described below in Sec.\,\ref{sec:results}. All electrodes and the magnetic field strength along the symmetry axis are depicted in Fig.\,\ref{fig:simulation_fieldstrength} with the detector at the end of the drift section.
The three-dimensional magnetic field was computed with a Helmholtz-coil model and fitted to measured values with the method of maximum-likelihood. The magnetic field strength at the position of the measurement trap is $B=\unit[7]{T}$.

\subsection{Monitoring of the magnetron motion}
\label{sec:simulation_dipolar}
\begin{figure}[htpb]
  \centering
  \subfigure[]{
    \label{fig:simulation_dipolar_radii_a}
    \includegraphics[width=0.45\textwidth]{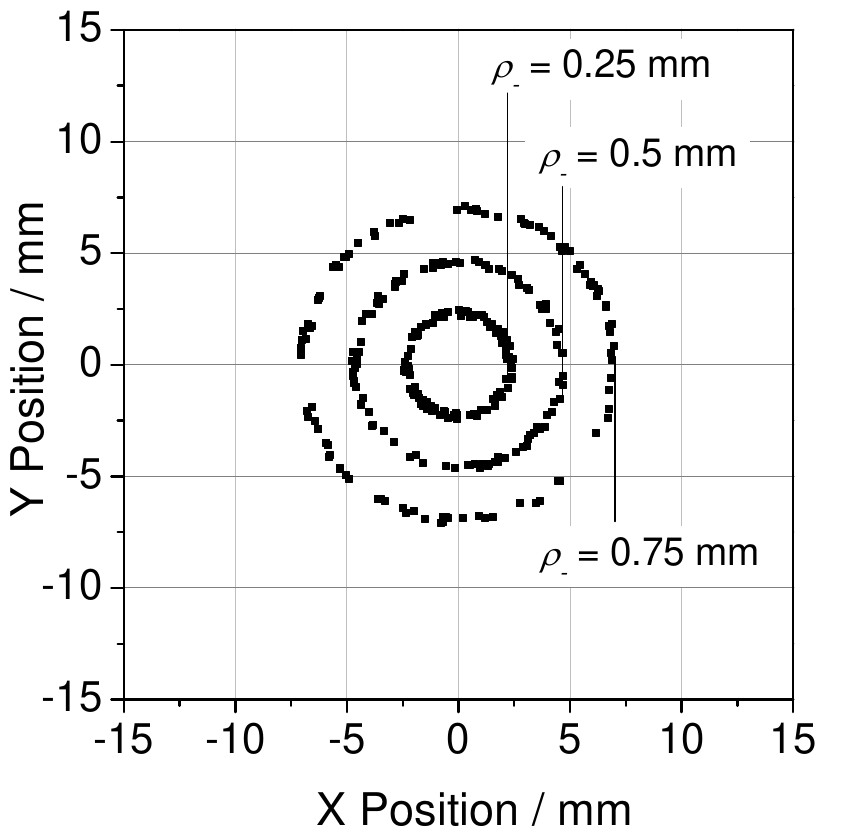}
  }
  \subfigure[]{
    \label{fig:simulation_dipolar_radii_b}
    \includegraphics[width=0.45\textwidth]{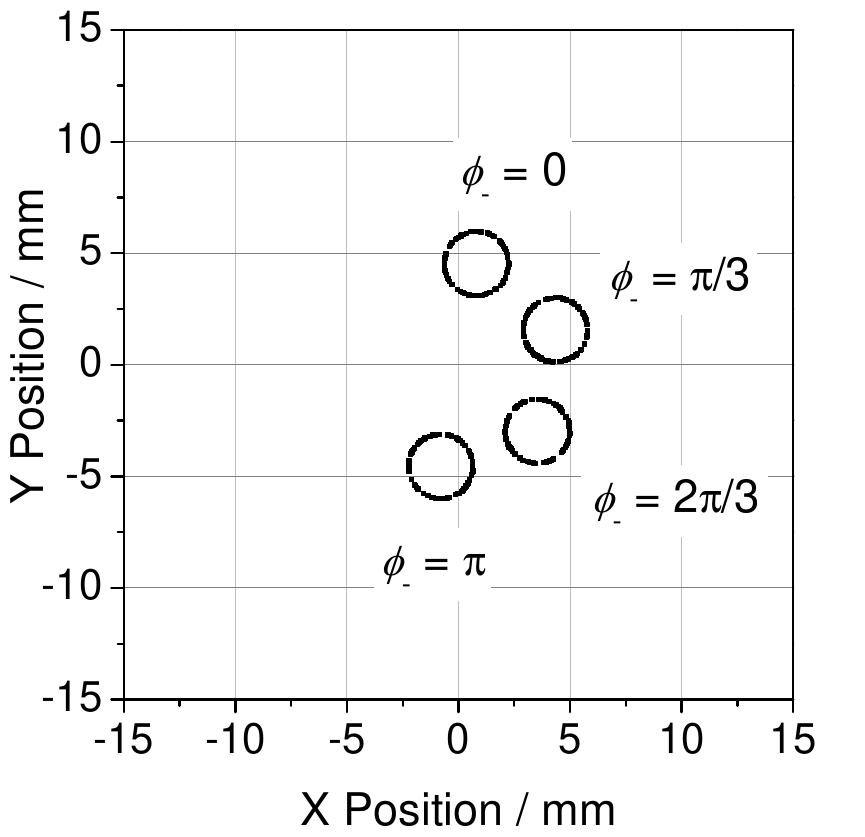}
  }
  \caption{Simulated detector images of ions with atomic mass $m=100$\,u ejected from a Penning trap after a dipolar excitation; (a) for three different magnetron radii (inner circle: $\rho_-=\unit[0.25]{mm}$, mid circle: $\rho_-=\unit[0.5]{mm}$, outer circle: $\rho_-=\unit[0.75]{mm}$) and a constant cyclotron amplitude $\rho_+=\unit[2.5]{\mu m}$; (b) for four values of the magnetron phase $\phi_-=(0,\pi/3,2\pi/3,\pi)$ with $\rho_-=\unit[0.5]{mm}$ and an increased cyclotron amplitude $\rho_+=\unit[25]{\mu m}$.}
\end{figure}

The imaging of the magnetron motion in the measurement trap on the position-sensitive detector was investigated for singly charged positive ions with an atomic mass of 100\,u. To this end, ions started with a fixed magnetron radius $\rho_-$ and were ejected to the detector after half a magnetron period. Since \mbox{$\rho_-\gg\rho_+$} was chosen, the ions possessed only little radial energy. Therefore, the drift voltages had to be adjusted in order to make the magnetron radius observable on the detector. For the simulation three magnetron radii each with 100 ions were considered and the phase of the magnetron motion $\phi_-$ was chosen randomly. The acquired picture in Fig.\,\ref{fig:simulation_dipolar_radii_a} shows three separated concentric rings with a radius~$R$ proportional to $\rho_-$ \mbox{($R=9.8\,\rho_-$)}. 
Thus, the final distance from the symmetry axis is a measure for the magnetron radius $\rho_-$ at the time of ejection.\newline
If the magnetron phase $\phi_-$ of all ions is set equal, the detected events are aggregated in a spot. This is demonstrated in  Fig.\,\ref{fig:simulation_dipolar_radii_b} for different values of the magnetron phase. Since the cyclotron radius $\rho_+$ was increased, the cyclotron motion is visible as small rings. An additional axial motion caused a small variation in time of flight but did not affect the detector pictures.\newline
The results show that the DLD40 could be used as a reliable tool to monitor the state of the ion motion before the ejection from the trap. The results were also confirmed by experimental studies as described below in Sec.\,\ref{sec:results}.

\subsection{Separation of ion species after quadrupolar excitation}
\label{sec:simulation_quadrupolar}
\begin{figure}[htpb]
  \centering
    \includegraphics[width=0.5\textwidth]{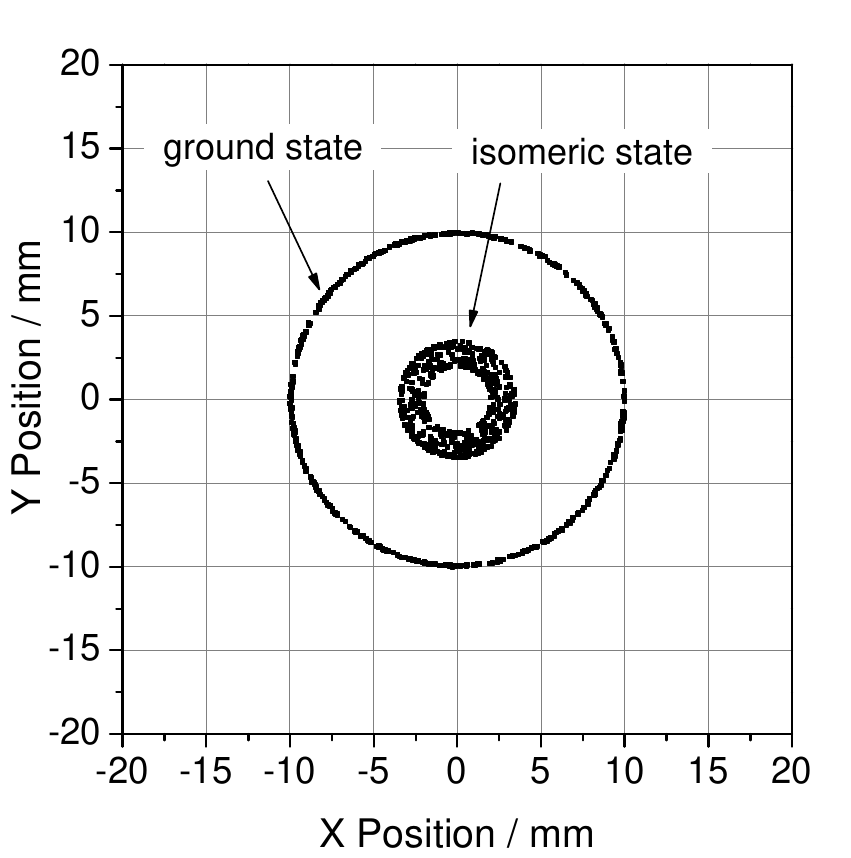}
    \caption{Simulation result: detector picture after a quadrupolar excitation of ions with mass $m_1$ and $m_2=m_1+\Delta m$ at a fixed excitation frequency $\nu_{rf}=\nu_{c,m_1}$. The outer ring is formed by resonantly excited ions with mass $m_1$ and the inner ring corresponds to the off-resonant ions in the isomeric state with mass $m_2$. The thickness of the inner ring is due to a slight cyclotron motion of the isomeric ions.}
    \label{fig:simulation_quadrupolar}
\end{figure}

\begin{figure}[htpb]
  \centering
  \subfigure[]{
    \label{fig:simulation_quadrupolar_TOF_a}
    \includegraphics[width=0.47\textwidth]{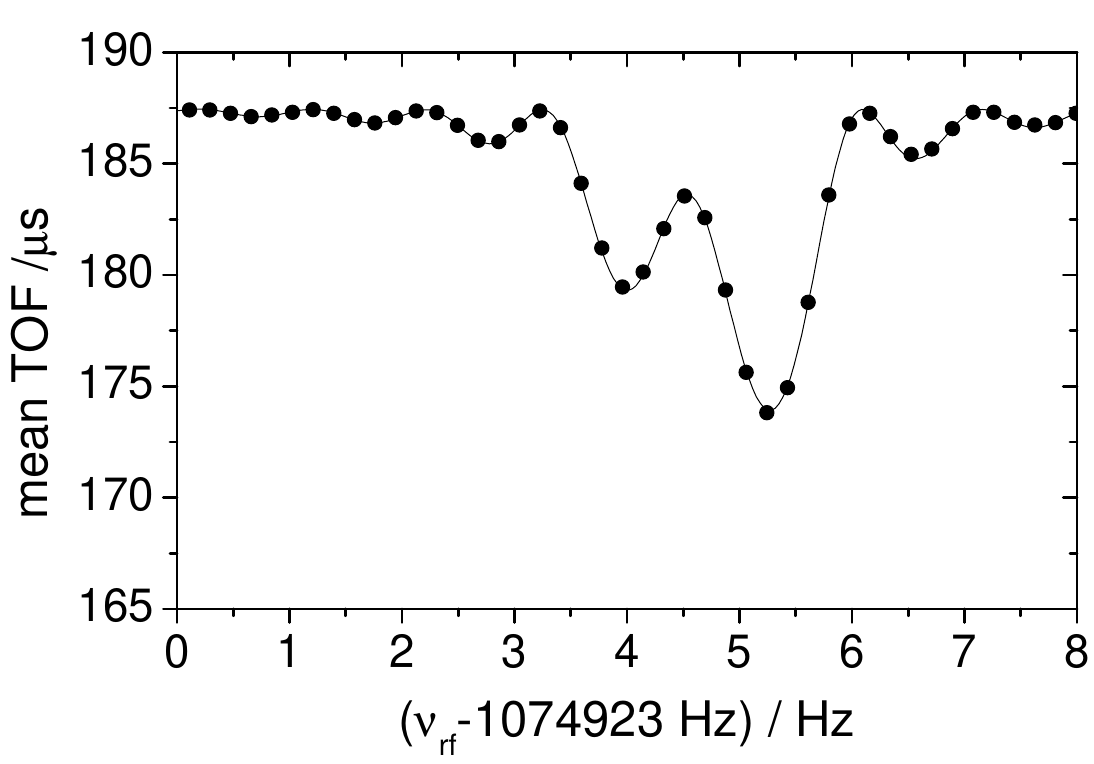}
  }
  \subfigure[]{
    \label{fig:simulation_quadrupolar_TOF_b}
    \includegraphics[width=0.47\textwidth]{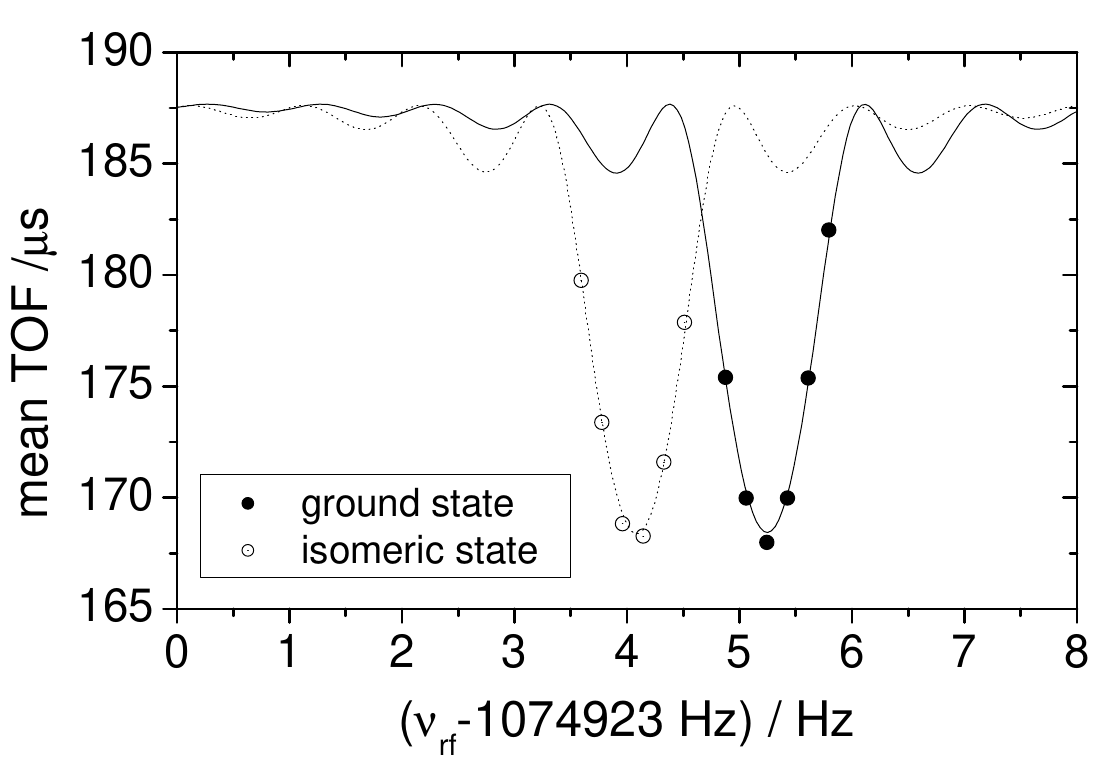}
  }
    \caption{Results of a simulated TOF-ICR experiment for hypothetical masses $m_1$ and $m_2$ with an abundance ratio of 2:1: (a) mean TOF-values and (b) individual TOF-values obtained via spatial separation in the region of resonance. The excitation time is $T_{rf}=\unit[1]{s}$. Fits of the theoretical line shape are given by solid and dotted lines, respectively.}
\end{figure}

In order to improve the contrast in a TOF-ICR measurement when contaminant ions are present in the trap, it was investigated whether the position of the ions on the detector provides information about the degree of conversion after the quadrupolar excitation. If this was the case, the DLD40 could be used to separate contaminant nuclei from the ions of interest. The principle of this separation method will be addressed in the following.\newline
Two ion species with masses $m_1$ and $m_2=m_1+\Delta m$ are considered, which have an initial magnetron radius $\rho_-$, neglecting any cyclotron amplitude. When an azimuthal quadrupolar rf-field at the cyclotron frequency $\nu_c=qB/(2\pi m_1)$ is applied, the resonantly excited ions perform a periodic conversion between pure magnetron motion and pure cyclotron motion, while those ions having mass $m_2$ only reach a lower degree of conversion~\cite{koen95}. Since $\omega_c\gg\omega_-$, the radial energy after an excitation is a measure for the degree of conversion. When the ions pass the drift section, the radial amplitude increases due to the magnetic field gradient. Since the resonantly excited ions reach a greater radius, different masses can be identified by the position-resolving detection method.\newline
This technique was implemented in a simulation for $m_1=100\,u$ and a hypothetical low-lying isomeric state with an excitation energy of $\unit[100]{keV}$ corresponding to a relative mass shift $\Delta m/m_1=1\cdot10^{-6}$. Both ion species start with a fixed magnetron phase and a random cyclotron phase. When the rf-frequency is set to $\nu_{rf}=\nu_{c,m_1}$, and the rf-amplitude $V_{rf}$ and the excitation duration $T_{rf}$ are chosen to accomplish a full conversion, the two ion species are successfully separated as shown in Fig.\,\ref{fig:simulation_quadrupolar}.
All ions with mass $m_1$ are located on an outer ring, whereas the isomeric species remain near to the symmetry axis. This method also works if the ions with mass $m_2$ are resonantly excited. Hence, all detected events can be assigned to one ion species in the case of resonant excitation. However, in the case of non-resonant excitation the separation is not so straight forward.\newline
In the following, the position-resolving detection is applied to a \mbox{TOF-ICR} experiment for $m_1$ and $m_2$. The TOF-contrast is defined as
\beq
 \textnormal{TOF-contrast}/\,\% =100\cdot(1-TOF_{min}/TOF_{base})\,,
 \label{eq:TOF_contrast}
\eeq
where $TOF_{base}$ is the mean time of flight far away from the resonance and $TOF_{min}$ is the one in resonance. First, a conventional experiment is simulated for an abundance ratio of ions in the ground and in the isomeric state, respectively, of 2:1. Usually, the two ion species cannot be distinguished and the TOF-contrast is lowered as shown in~Fig.\,\ref{fig:simulation_quadrupolar_TOF_a}. 
When position-sensitive detection is employed, the time of flight of each isomeric state can be determined individually in the region of resonance. This is shown in Fig.\,\ref{fig:simulation_quadrupolar_TOF_b} where fits of the theoretical line-shape were applied to the respective set of data points. The TOF-contrast is raised from 6.1\,$\%$ to 10.4\,$\%$ for $m_1$ and from 3.5\,$\%$ to 10.4\,$\%$ for $m_2$, respectively. The presented method could be applied in high-precision Penning trap mass spectrometry and might help to improve the precision in the determination of the cyclotron frequency of an ion of interest in case of the presence of a contaminating ion species. Frequency shifts caused by ion-ion interactions~\cite{boll92} are of course not affected by this method. However, since the presented technique aims at experiments with a very small number of ions, these effects are of no concern.
Experimental results are discussed in Sec.\,\ref{sec:results_quadrupolar}.

\section{Experimental studies}
\label{sec:results}
\begin{figure}[htpb]
  \centering
   \includegraphics[width=\textwidth]{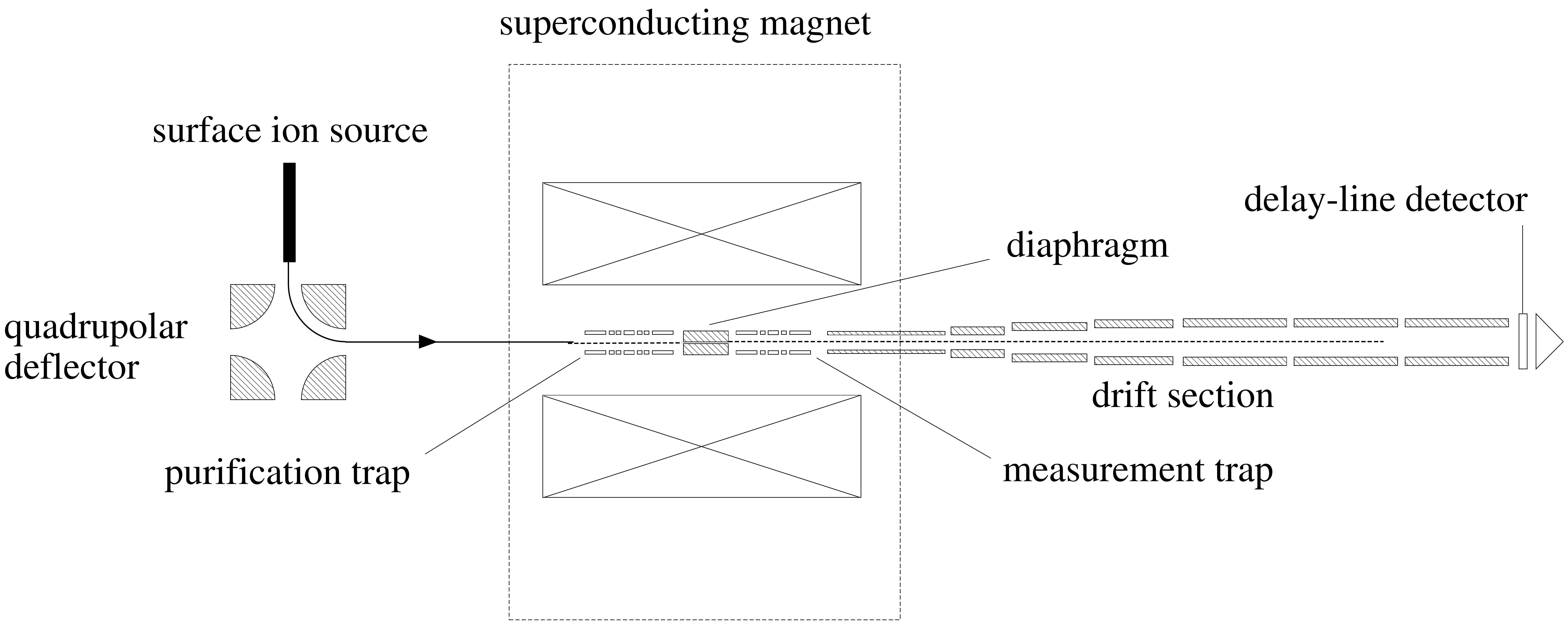}
    \caption{Sketch of the experimental setup of the SHIPTRAP mass spectrometer.}
    \label{fig:results_SHIPTRAP}
\end{figure}

In order to confirm the simulation results described above, experimental studies were performed at the SHIPTRAP facility~\cite{bloc05} located at the GSI Helmholtz Centre for Heavy Ion Research in Darmstadt. SHIPTRAP is a double Penning trap mass spectrometer connected to the velocity filter SHIP~(Separator for Heavy Ion reaction Products)~\cite{muen79,hofm00}, where superheavy elements produced in fusion reactions are separated from an ion-beam. SHIPTRAP is composed of two cylindrical Penning traps in the bore of a 7-T superconducting magnet (see Fig.\,\ref{fig:results_SHIPTRAP}).
The first trap is used for the preparation of the incoming ion beam employing mass selective buffer gas cooling~\cite{sava91}. In the second trap the cyclotron-frequency measurement is performed using the TOF-ICR technique. The magnetic field strength along the $z$-axis and the arrangement of the ion optics behind the measurement trap are shown in Fig.\,\ref{fig:simulation_fieldstrength}. As described in Sec.\,\ref{sec:simulation}, the DLD40 detector was mounted at the end of the drift section, where the magnetic field strength is weak enough ($B<\unit[10]{mT}$) in order not to affect the detection process. In the measurements described in this work, SHIPTRAP was operated off-line, with a surface ion-source providing $^{133}$Cs$^+$ and $^{85,87}$Rb$^+$ ions~\cite{kirchner1987isb}.\newline
In accordance with the simulation the experimental study is twofold. First, the imaging of the magnetron motion is realized for $^{133}$Cs ions. Second, the separation of contaminating ions is probed for $^{85,87}$Rb. Each measurement requires a specific setting of the ion optics due to the different radial energy of the detected ions. When the radial energy is increased by means of the conversion from magnetron to cyclotron motion, the ion optics have to be adjusted for a strong focusing and thus the magnetron motion is not observable anymore. To obtain a difference in time of flight a slowdown of the ion transport in the region of the highest magnetic field gradient is also needed. The corresponding settings of the drift voltages are given in Tab.\,\ref{tab:results_voltages}.
\begin{table}[htpb]
   \centering
       \begin{tabular}{|l||r|r|r|r|r|r|r|r|}
           \hline
           Electrode &  Out1 & Out2 & Out3 & Exit1 & Exit2 & Exit 3 & Exit 4 & Exit5\\
           \hline
           \hline
           \scriptsize{\textbf{Voltage a)}} & \scriptsize{-100\,{V}} & \scriptsize{-150\,{V}} & \scriptsize{-200\,{V}} & \scriptsize{-300\,{V}} & \scriptsize{-1\,{kV}} & \scriptsize{-1.3\,{kV}} & \scriptsize{-1{kV}} & \scriptsize{-1.1\,{kV}} \\
           \hline
           \scriptsize{\textbf{Voltage b)}} & \scriptsize{-40\,{V}} & \scriptsize{-3\,{V}} & \scriptsize{-40\,{V}} & \scriptsize{-350\,{V}} & \scriptsize{-1.1\,{kV}} & \scriptsize{-960\,{V}} & \scriptsize{-1.1\,{kV}} & \scriptsize{-400\,{V}} \\
           \hline
       \end{tabular}
   \caption[Focusing voltages for dipolar excitation]{Voltages at the drift electrodes of SHIPTRAP used for the monitoring of the magnetron motion (a), and for the separation of impurities (b). The reasons for using two sets of voltages are explained in the text.}
   \label{tab:results_voltages}
\end{table}

\subsection{Monitoring of the magnetron motion}
\label{sec:results_dipolar}
\begin{figure}[htpb]
  \centering
    \includegraphics[width=0.7\textwidth]{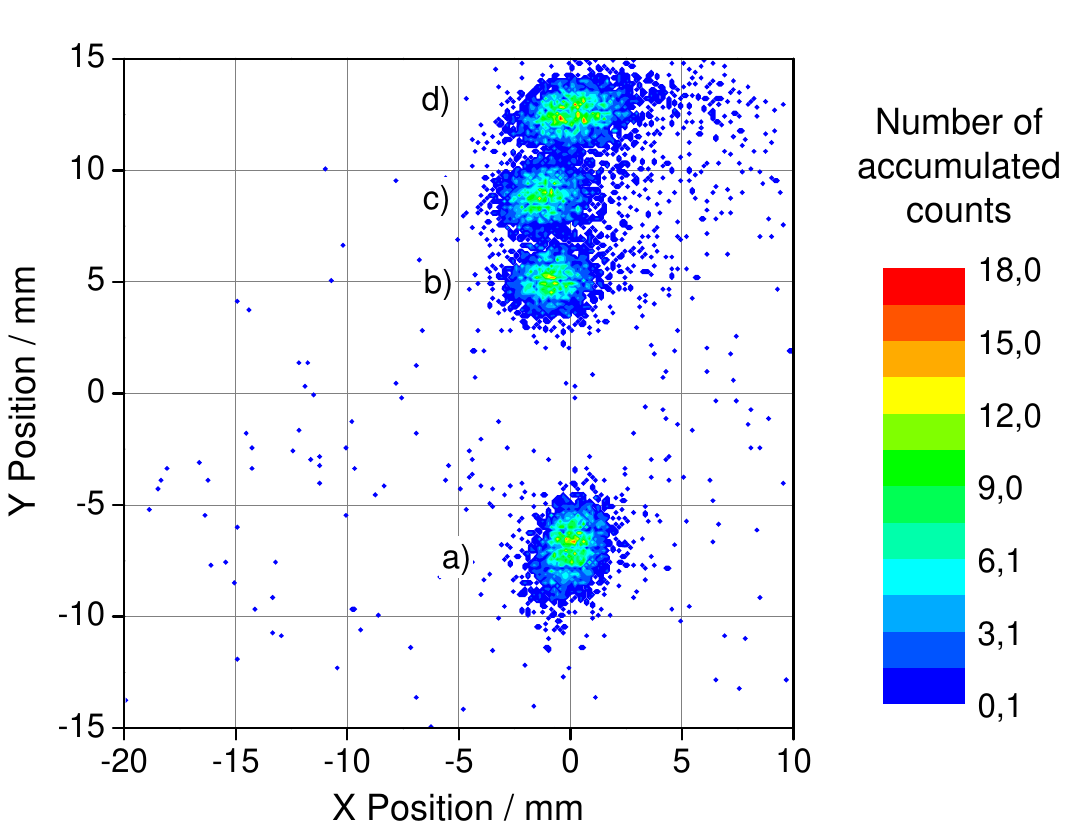}
    \caption{Experimental data obtained for different magnetron excitation durations and a constant trapping time: (a)~$T_d=0$\,ms, (b)~$T_d=50$\,ms, (c)~$T_d=70$\,ms, and (d)~$T_d=90$\,ms. At the beginning of the excitation the ion cloud was confined in the trap center without radial motion corresponding to the lowermost spot. For clarity the detector data is shown with reduced resolution and not all recorded spots are included. The detector center is at ($x=0$, $y=0$).}
    \label{fig:results_dipolar_radii}
\end{figure}

\begin{figure}[htpb]
  \centering
    \includegraphics[width=0.55\textwidth]{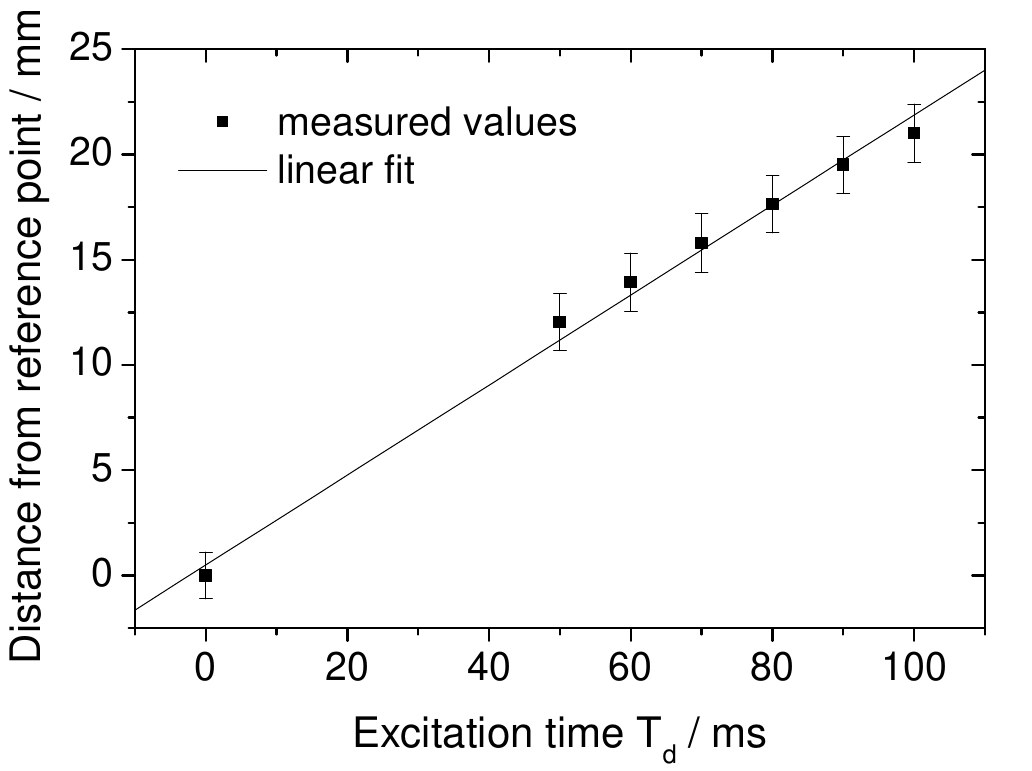}
    \caption{The displacement of all recorded spots plotted against the excitation duration $T_d$ with respect to the position of the lowermost spot in Fig.\,\ref{fig:results_dipolar_radii}. The linear fit evidences a linear dependence of the magnetron radius $\rho_-$ on the excitation duration $T_d$.}
        \label{fig:results_dipolar_radii_linear}
\end{figure}

\begin{figure}[htpb]
  \centering
    \includegraphics[width=0.7\textwidth]{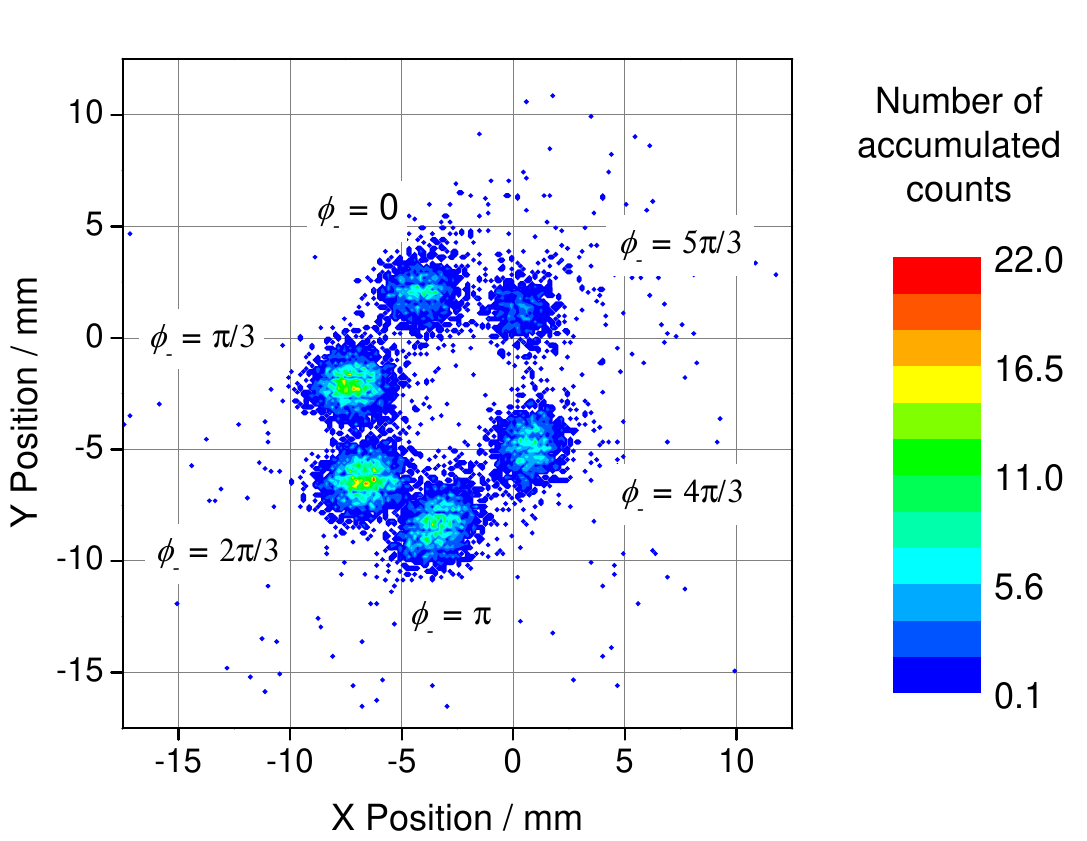}
    \caption{Experimentally recorded image of ejected ions for six magnetron phases $\phi_-$. The ejection of the ions was delayed for different time intervals in order to vary the magnetron phase at the time of ejection. The period of one magnetron motion for $^{133}$Cs in the used setup was $T_-=\unit[738]{\mu s}$.}
    \label{fig:results_dipolar_phase}
\end{figure}

First measurements were conducted concerning the observation of the magnetron motion as described in Sec.\,\ref{sec:simulation_dipolar}. \mbox{$^{133}$Cs$^+$ ions} were confined in the center of the measurement trap and excited by a phase-locked dipolar rf-field at the magnetron frequency $\nu_d=\nu_-=1355$\,Hz. The amplitude of the rf-field was $V_d=0.4$\,V and the excitation duration $T_d$ was set to several values between 0 and 100\,ms to observe different magnetron radii. To keep the overall trapping time constant an additional waiting time of $T_{wait}=\unit[100]{ms}-T_d$ was introduced. Hence, the magnetron phase at the time of ejection was unchanged. The data acquisition of the DLD40 was triggered by the ion ejection with a measurement time window of $\unit[200]{\mu s}$ and the position of each ion was recorded. This cycle was repeated once a second for 20\,minutes with in average two ions per shot. The data obtained for different excitation times $T_d$ is partly shown in Fig.\,\ref{fig:results_dipolar_radii}.
The detector picture shows Gaussian-shaped spots with a distinct position roughly along one line. This behavior was expected since the magnetron phase at the time of ejection was not changed. The displacement of a spot depends linearly on the magnitude of $T_d$, which is shown in Fig.\,\ref{fig:results_dipolar_radii_linear}.\newline
In a second measurement, snapshots of an ion cloud moving with a fixed magnet\-ron radius were recorded.
Therefore, the excitation time was fixed to \mbox{$T_d=\unit[50]{ms}$} to excite all ions to the same magnetron amplitude. By changing the waiting time interval after the magnetron excitation $T_{wait}$, the magnetron phase $\phi_d$ was evolved up to one period of motion. From the simulation studies, ion spots at several positions on a circular orbit were expected. This was confirmed in the experiment as shown in Fig.\,\ref{fig:results_dipolar_phase}.
However, the arrangement of spots differed slightly from a perfect circle and the count rate was changing depending on the magnetron phase. Since the detector efficiency is homogeneous over the surface, this indicates a small asymmetry in the electromagnetic field between the Penning trap and the ion detector.\newline
The presented results evidence that a detailed control of the state of magnetron motion is possible using the DLD40. Using the proportionality determined in Sec.\,\ref{sec:results_dipolar} and the average full width at half maximum of the detected spots (${FWHM}=1.88$\,mm) one obtains a resolution for the magnetron radius of $\Delta \rho_-=\unit[0.19]{mm}$. At a magnetron radius of $\unit[1.2]{mm}$ the magnetron phase can be determined with a resolution of $\Delta \phi_-= 8^\circ$.

\subsection{Separation of Rb isotopes}
\label{sec:results_quadrupolar}
\begin{figure}[htpb]
  \centering
    \includegraphics[width=0.7\textwidth]{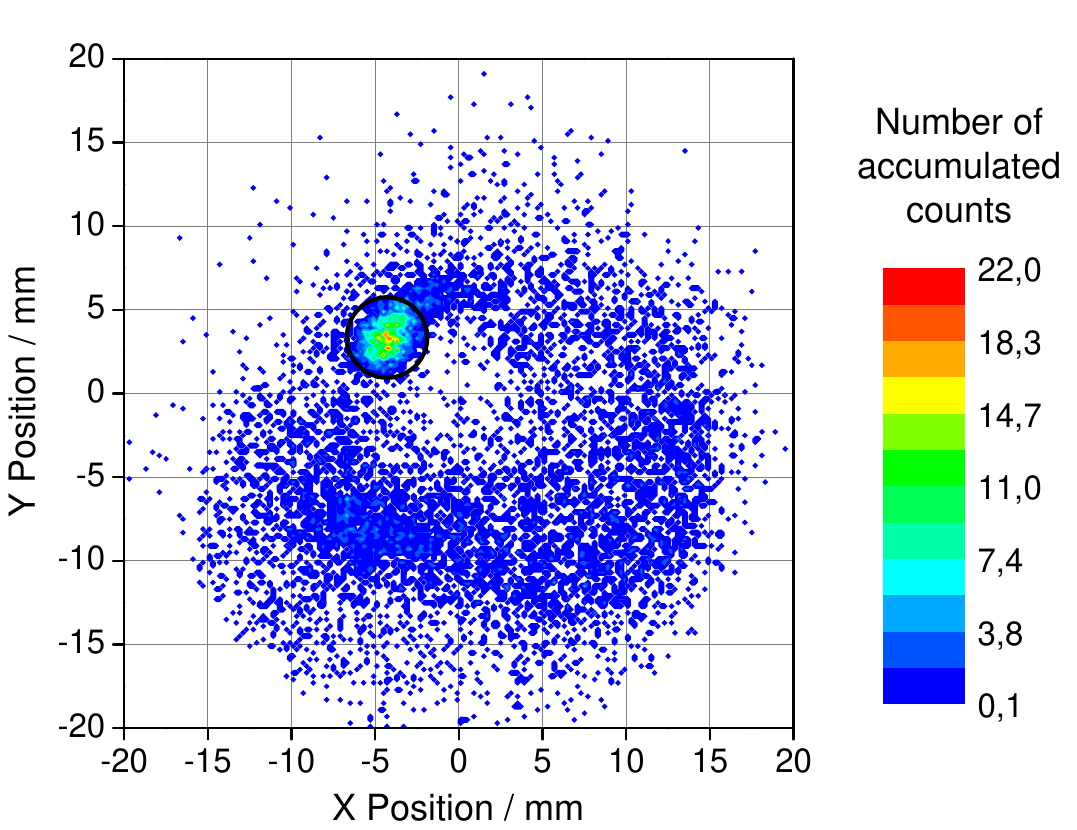}
    \caption{Experimental result of a quadrupolar excitation of $^{85}$Rb$^+$ and $^{87}$Rb$^+$ at the cyclotron frequency of $^{85}$Rb$^+$. The spot enclosed by a black circle corresponds to $^{87}$Rb ions having little radial energy due to the non-resonant excitation.}
    \label{fig:results_quadrupolar_a}
\end{figure}

\begin{figure}[htpb]
  \centering
  \subfigure[]{
    \label{fig:results_quadrupolar_TOFa}
    \includegraphics[width=0.47\textwidth]{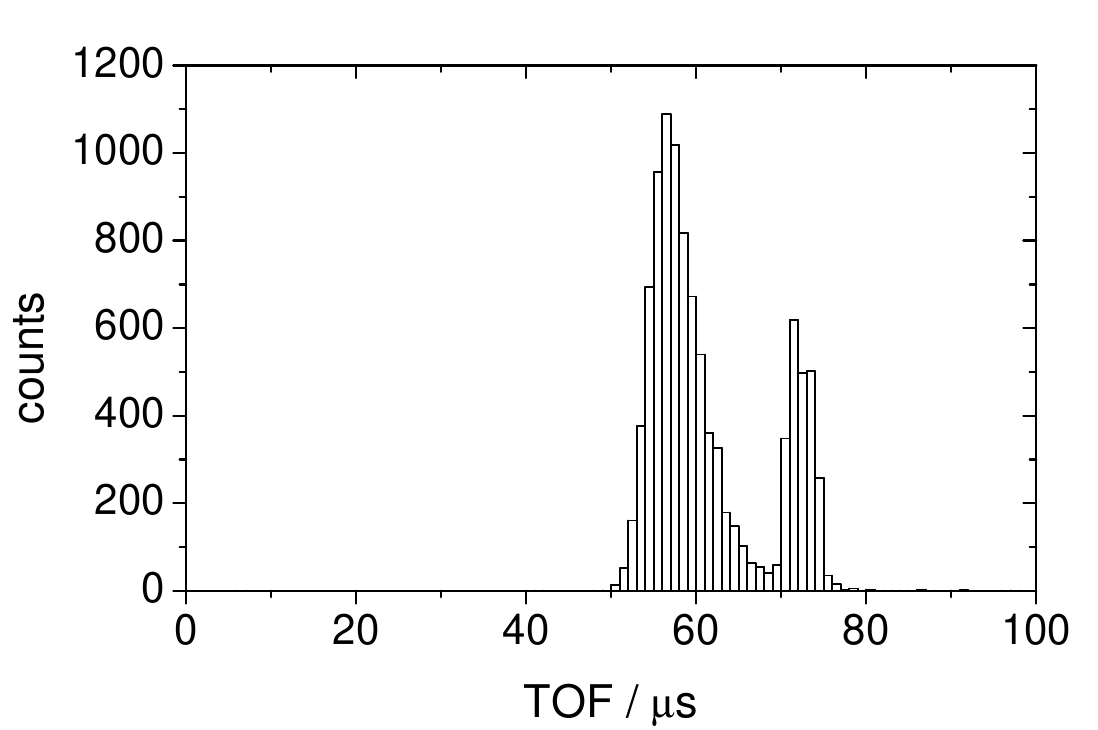}
  }
  \subfigure[]{
    \label{fig:results_quadrupolar_TOFb}
    \includegraphics[width=0.47\textwidth]{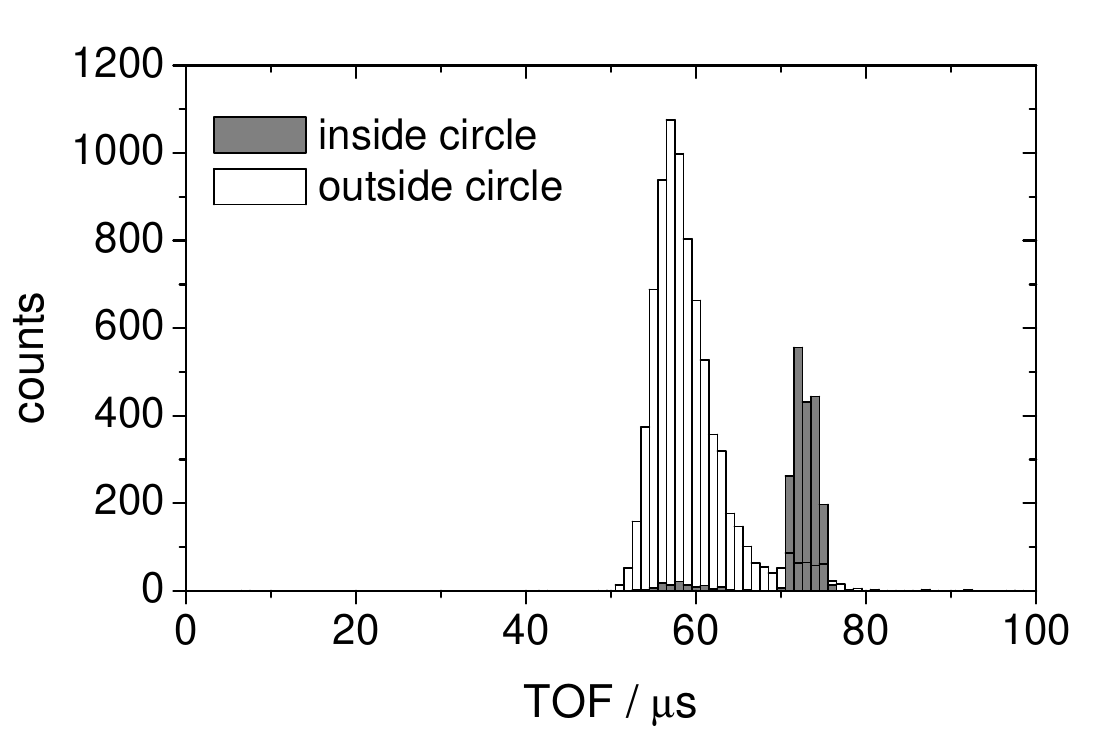}
  }
    \caption{Time-of-flight analysis for Fig.\,\ref{fig:results_quadrupolar_a}: (a) overall \mbox{time-of-flight} distribution and (b) \mbox{time-of-flight} distributions for the ions inside (gray bars) and outside (white bars) the marked area. The peak at $TOF=\unit[58]{\mu s}$ corresponds to $^{85}$Rb$^+$ and the natural abundance ratio of $^{85,87}$Rb of 3:1 explains the difference in intensity of the two peaks.}
\end{figure}

\begin{figure}[htpb]
  \centering

  \subfigure[]{
    \label{fig:results_resonance_TOFa}
    \includegraphics[width=0.47\textwidth]{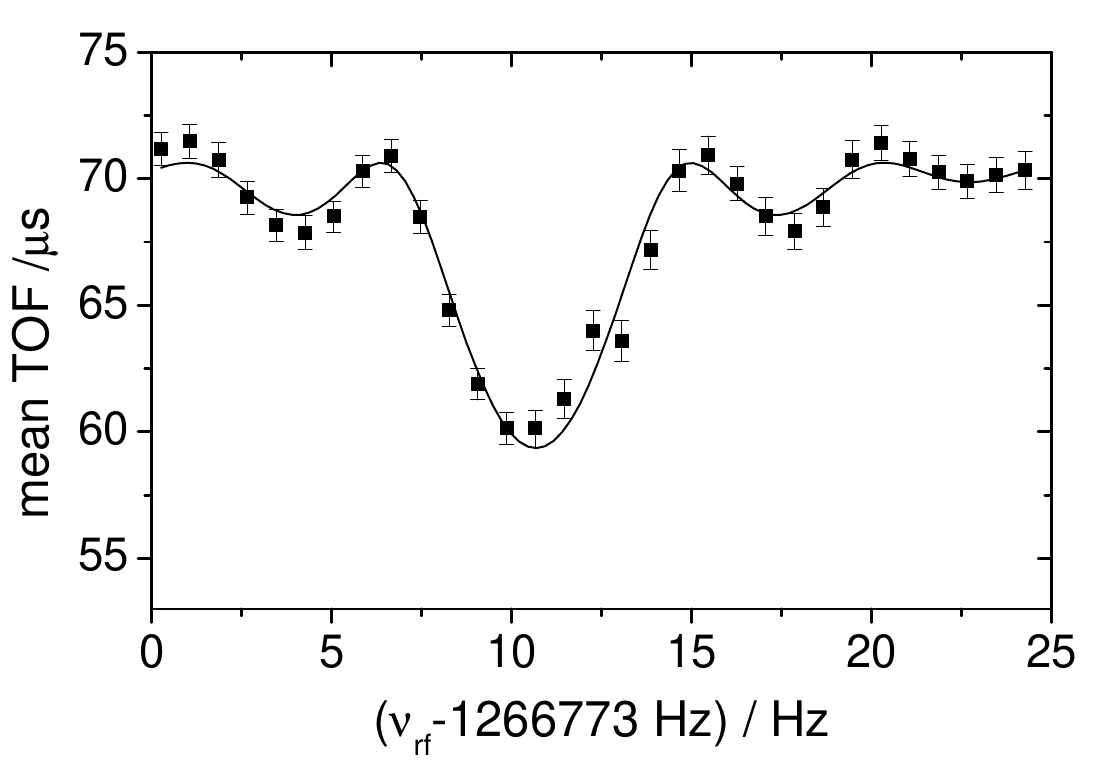}
  }
  \subfigure[]{
    \label{fig:results_resonance_TOFb}
    \includegraphics[width=0.47\textwidth]{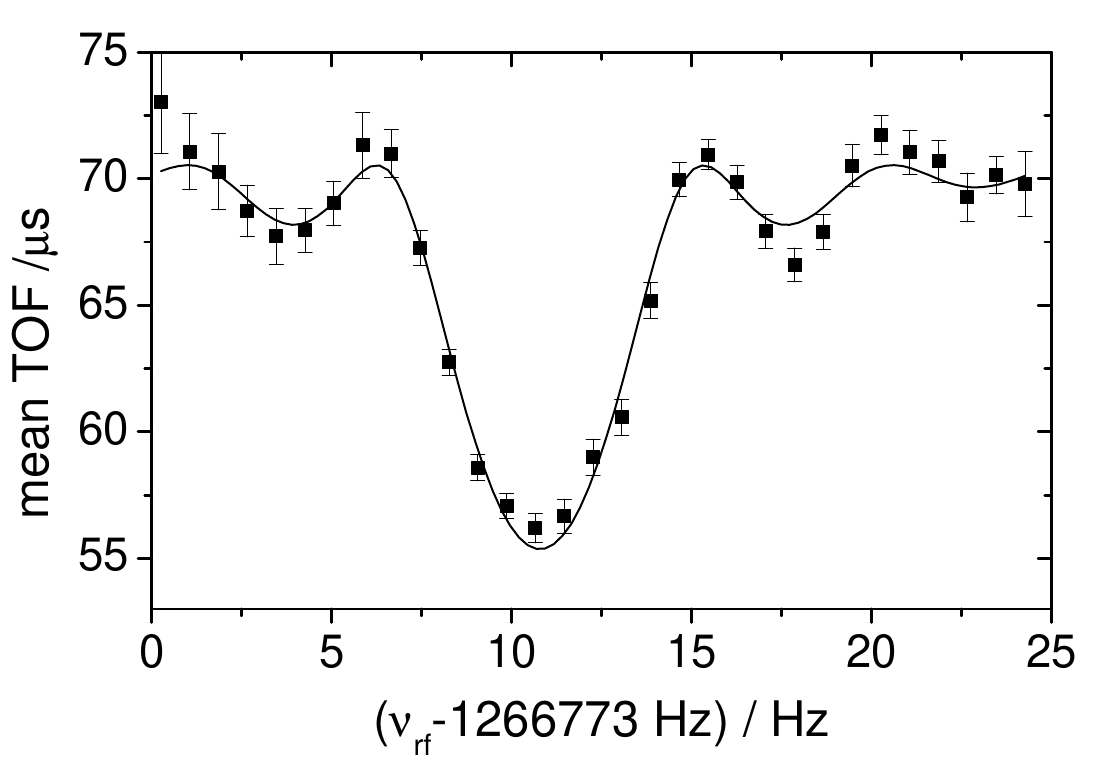}
  }

    \caption{Time-of-flight resonance for a mixture of $^{85}$Rb and $^{87}$Rb ions: (a) for all recorded events and (b) for those events selected via position resolving detection. The solid line is a fit of the theoretical line shape to the data points by which the cyclotron frequency is determined~\cite{koen95}.}
\end{figure}

\begin{figure}[htpb]
  \centering
    \includegraphics[width=0.7\textwidth]{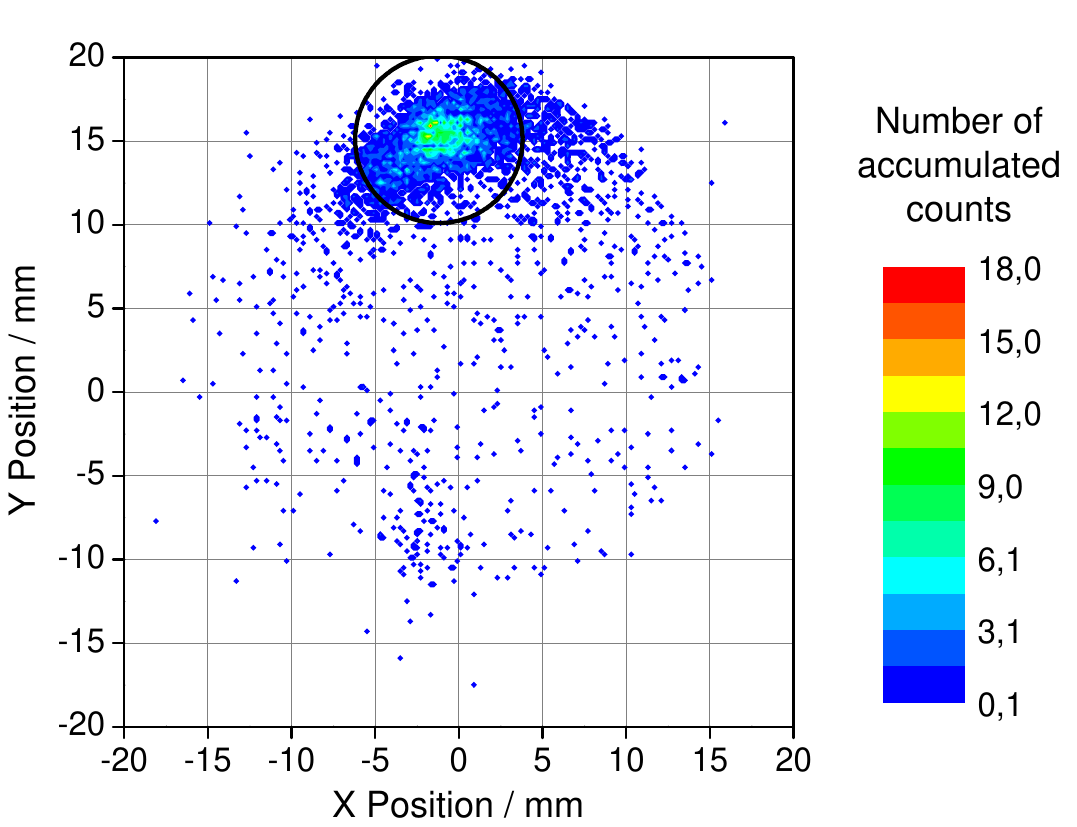}
    \caption{Detector events recorded in a TOF-ICR measurement of a mixture of $^{85}$Rb and $^{87}$Rb ions. The black circle encloses events neglected in the time-of-flight analysis of the resonance shown in Fig.\,\ref{fig:results_resonance_TOFb}.}
    \label{fig:results_resonance}
\end{figure}

In accordance with the numerical studies a quadrupolar excitation was applied on $^{85}$Rb$^+$ and $^{87}$Rb$^+$ stored simultaneously in the measurement trap. The excitation scheme is as follows. At first, both ion species were excited to the same magnetron radius by a dipolar excitation at the magnetron frequency ($\nu_d=1337$\,Hz, $V_d=0.4$\,V, $T_d=50$\,ms). In a second step, the ions were subjected to a quadrupolar rf-field at the true cyclotron frequency of $^{85}$Rb$^+$ ($\nu_{rf}=1.266$\,MHz, $V_{rf}=6.4$\,V, $T_{rf}=100$\,ms) resulting in a periodical conversion from magnetron motion into cyclotron motion for the $^{85}$Rb$^+$ ions. Subsequently, all ions were ejected from the measurement trap.\newline
In the corresponding detector picture a spot with a high amount of events is visible at ($x=-5$\,mm, $y=4$\,mm) in a background of wide spread single events (see Fig.\,\ref{fig:results_quadrupolar_a}).
The distortion from the expected circular shape (see Fig.\,\ref{fig:simulation_quadrupolar}) is likely due to a slight misalignment of the ion transport. Additionally, the non-perfect shape of the ejection pulse was not considered in the simulation and could affect the ion trajectories. From the computational results one expects the ions in the spot to be $^{87}$Rb. This is evidenced by a \mbox{time-of-flight} analysis. The \mbox{time-of-flight} distribution of the whole detector area shows two clearly separated peaks in Fig.\,\ref{fig:results_quadrupolar_TOFa} with a weighting corresponding to the natural abundance of $^{85,87}$Rb \mbox{($^{85}$Rb: 72.2\,$\%$, $^{87}$Rb: 27.8\,$\%$)}.
Thus, events with a flight time greater than $\unit[70]{\mu s}$ can certainly be assigned to $^{87}$Rb. If the time of flight is determined for the areas inside and outside a rectangular area enclosing the aforementioned spot, it appears that nearly all ions having a flight time longer than $\unit[70]{\mu s}$ are accumulated in the spot \mbox{(see Fig.\,\ref{fig:results_quadrupolar_TOFb})}. Therefore, it is evidenced that $^{85}$Rb and $^{87}$Rb were spatially separated in agreement with the simulations.\newline
In the last experimental test a full resonance was recorded to demonstrate the application of position resolving detection in TOF-ICR measurements. Again a mixture of $^{85}$Rb and $^{87}$Rb with an abundance ratio of 3:1 was probed in the measurement trap. A repeated scan of the excitation frequency $\nu_{rf}$ around the cyclotron frequency of $^{85}$Rb leads to the resonance curve shown in Fig.\,\ref{fig:results_resonance_TOFa} where the TOF-contrast is determined to be 16\,\% according to~Eq.\,(\ref{eq:TOF_contrast}).
The corresponding distribution of events is shown in Fig.\,\ref{fig:results_resonance} where those ions having low radial energy are located around (x=-2\,mm, y=16\,mm).
If all events lying closer than 5\,mm to this position are neglected in computing the mean time of flight, the TOF-contrast is raised to $22$\,$\%$. Even though the statistics for $^{85}$Rb have been decreased due to this procedure by 50\,$\%$, however mainly in the baseline, the uncertainty in determining the cyclotron frequency is decreased by almost $40$\,$\%$ (see Fig.\,\ref{fig:results_resonance_TOFb}).\newline
The presented results prove that the principle separation scheme works for two well-known ion species having a large difference in cyclotron frequency of \mbox{$\Delta\nu_c\approx29$\,kHz}. However, simulation results indicate that also masses not resolved in \mbox{time-of-flight} measurements could be spatially separated using the described technique. This still has to be proved in an experiment, preferably with low lying isomeric states or with close-by molecular contaminant species. 

\section{Outlook}
\label{sec:Outlook}
In the presented work, the position-sensitive ion detection technique was introduced for the use in Penning trap mass spectrometry experiments. By using the DLD40, the position of the ejected ions can be recorded precisely and single events can be localized with an uncertainty of less than $\unit[0.1]{mm}$. The position information helps optimizing the ion optics as well as the excitation parameters. Previously, only the ion count rate and time of flight used to be recorded. It was shown in computational simulations that the DLD40, used in combination with a common TOF-ICR Penning trap setup, offers a tool for accurate observation of the radial ion motion. These results were confirmed in experimental tests, in which a variety of magnetron radii and phases of the magnetron motion were successfully resolved. In this way a full control of the amplitude and the phase of the magnetron motion was demonstrated. This technique allows for a faster adjustment of the experimental setup since the magnetron radius is usually deduced from a full cyclotron resonance by a fit of the theoretical line-shape.\newline
Furthermore, the results demonstrate that contaminants in the measurement trap could be separated from the ions of interest using position-resolving detection. This leads to an improved TOF-contrast which has been demonstrated successfully for the case of $^{85}$Rb$^+$ with $^{87}$Rb$^+$ as a contamination. The presented computational results show that this is also possible for a pair of nuclei with a difference in cyclotron frequency below the Fourier limit of resolution. In future \mbox{TOF-ICR} experiments, this technique could be employed to improve the resonance strength by neglecting contributions to the mean time of flight coming from impurities. In the near future investigations with the new detection technique will take place at high precision Penning trap mass spectrometers at SHIPTRAP~\cite{bloc05} at the GSI Helmholtz Centre for Heavy Ion Research in Darmstadt and at \mbox{TRIGA-TRAP}~\cite{ketelaer08} at the Institute for Nuclear Chemistry of the University of Mainz. Furthermore, an update of the DLD40, capable of strong magnetic fields, has recently been tested at the Max-Planck-Institute for Nuclear Physics in Heidelberg.

\section{Acknowledgement}
\label{sec:Ack}
The support of the Helmholtz Association of National Research Centres (HGF) under contract number VH-NG-037, of the Deutsche Forschungsgemeinschaft (BL981-2-1) and of the German Federal Ministry for Education and Research (06MZ215) is gratefully acknowledged. We also acknowledge the support of Horst Schmidt-B\"ocking and the RoentDek company. Sz.~Nagy acknowledges support from the Alliance Program of the Helmholtz Association (HA216/EMMI).




%
%

\bibliographystyle{elsarticle-num}

\bibliography{bibliography3.bib}

\end{document}